\documentclass[12pt,a4paper]{article}
\usepackage{amsmath}
\usepackage{amsfonts}
\usepackage[english]{babel}
\usepackage{amssymb}
\usepackage{graphicx}
\usepackage{hyperref}
\usepackage[left=2cm,right=2cm,top=2cm,bottom=2cm]{geometry}

\usepackage{xcolor}

\newcommand{\be}{\begin{equation}}
\newcommand{\ee}{\end{equation}}
\newcommand{\bea}{\begin{eqnarray}}
\newcommand{\eea}{\end{eqnarray}}

\newtheorem{remark}{Remark}

\title{Dynamical aspects in the Quantizer-Dequantizer formalism}
\author{Florio M. Ciaglia, Fabio Di Cosmo\\ 
\textit{Dipartimento di Fisica, Universit\`a di Napoli ``Federico II"}\\
\textit{Via Cinthia Edificio 6, I-80126 Napoli, Italy}\\
\textit{and INFN-Sezione di Napoli, Via Cinthia Edificio 6, I-80126 Napoli, Italy}\\ 
\and Alberto Ibort\\
\textit{Departamento de Matem\'{a}ticas, Universidad Carlos III de Madrid,}\\
\textit{ Avda. de la Universidad 30, 28911 Legan\'{e}s, Madrid, Spain.}\\
\textit{ICMAT, Instituto de Ciencias Matem\'{a}ticas (CSIC - UAM - UC3M - UCM),}\\
\textit{ Nicol\'{a}s Cabrera,13--15, Campus de Cantoblanco, UAM, 28049, Madrid, Spain.}\\
\and Giuseppe Marmo\\ 
\textit{Dipartimento di Fisica, Universit\`a di Napoli ``Federico II"}\\
\textit{Via Cinthia Edificio 6, I-80126 Napoli, Italy}\\
\textit{and INFN-Sezione di Napoli, Via Cinthia Edificio 6, I-80126 Napoli, Italy}
}
\date{}

\begin{document}

\maketitle

\begin{abstract}
The use of the quantizer-dequantizer formalism to describe the evolution of a quantum system is reconsidered.
We show that it is possible to embed a manifold in the space of quantum states of a given auxiliary system by means of an appropriate quantizer-dequantizer system. 
If this manifold of states is invariant with respect to some unitary evolution, the quantizer-dequantizer system provides a classical-like realization of such dynamics, which in general is non linear. 
Integrability properties are also discussed.
Weyl systems and generalized coherente states are used as a simple illustration of these ideas.
\end{abstract}

\section{Introduction} \label{section: introduction}

Interference phenomena are ubiquitous in Quantum Mechanics and this led Dirac to state that the (linear) superposition principle is one of the main features of a Quantum description of Physics \cite{dirac-principles_of_quantum_mechanics}. 
Indeed, the mathematics of Quantum Mechanics entails linear structures in the Hilbert space $\mathcal{H}$ of the system, in the dynamical evolution given by Schr\"{o}dinger equation, and in the set of linear operators on $\mathcal{H}$. 
On the other hand, the degrees of freedom of a classical physical system are generally modelled on non-linear manifolds, and the dynamical evolution needs not to present any superposition rule.

Consequently, if we want to understand a suitable classical limit of Quantum Mechanics we face the necessity of introducing nonlinear coordinates transformations in the formulation of Quantum Mechanics.
Indeed, this is precisely the motivation for the geometrical formulation of Quantum Mechanics  \cite{ashtekar_schilling-geometrical_formulation_of_quantum_mechanics,carinena_clemente-gallardo_marmo-geometrization_of_quantum_mechanics}.
An example of such a classical limit procedure is given by the so-called WKB short-wave limit of Schr\"{o}dinger equation which we now briefly recall.

Let $\psi(x,t)$ be a wavefunction in the Hilbert space $\mathcal{H}=\mathcal{L}^{2}(\mathbb{R}^{n}\,,\mathrm{d}\mu)$, where $\mathrm{d}\mu$ is the Lebesgue measure on $\mathbb{R}^{n}$, and consider the Schr\"{o}dinger equation:
\be\label{Schrodinger}
\imath\hbar\,\frac{\partial\,\psi}{\partial t}=-\frac{\hbar^{2}}{2m}\Delta\psi + V(x)\psi(x\,,t)
\ee
Using a polar representation $\psi(x\,,t)=A(x\,,t)\,\mathrm{e}^{-\frac{\imath}{\hbar} W(x\, ,t)}$, where the functions $A$ and $W$ are real, and $A$ is strictly positive, the Schr\"{o}dinger equation (\ref{Schrodinger}) becomes a system of two coupled partial differential equations:

\be\label{eqn: Q-H-J}
\frac{\partial W}{\partial t}=\frac{1}{2m}|\nabla W|^{2} + V(x) - \frac{\hbar^{2}}{2m}\frac{\Delta A}{A}\,,
\ee

\be\label{eqn pseudo-continuity equation}
\frac{\partial A}{\partial t}=\frac{1}{2m}\left( 2\nabla A\cdot \nabla W + A\Delta W\right)\,.
\ee
We would like to stress that this is a nonlinear change of coordinates in the Hilbert space which makes the superposition rule quite nontrivial.

Now, we may perform what is known as the classical limit of the Schr\"{o}dinger equation, which amounts to take the limit in which $\hbar$ goes to $0$.
It is clear that the only term which is affected by this limiting procedure is the third one in the right hand side of equation (\ref{eqn: Q-H-J}).
Clearly, when $\frac{\Delta A}{A}$ is bounded, if $\hbar$ goes to $0$, so does $\frac{\hbar^{2}}{2m}\frac{\Delta A}{A}$, and thus we can neglect the third term in equation (\ref{eqn: Q-H-J}).
The result is that the system of equations is no longer coupled, and Eq. (\ref{eqn: Q-H-J}) reduces to the Hamilton-Jacobi equation:

\be\label{eqn: H-J}
\frac{\partial W}{\partial t}=\frac{1}{2m}|\nabla W|^{2} + V(x) \, ,
\ee
associated with the Hamiltonian function $H=\frac{1}{2m}p^2 + V(x)$.

\begin{remark}
In the case in which $A=0$ at some isolated points, that is, the wavefunction $\psi$ has nodes, the third term of equation (\ref{eqn: Q-H-J}) could present divergences, and thus one should previously check that $\frac{\Delta A}{A}$ presents no divergences on the nodes, and then proceed as illustrated above.
If $\frac{\Delta A}{A}$ actually presents divergences, than the procedure outlined can not be applied to the quantum state described by the wavefunction $\psi$.
\end{remark}

Since we are dealing with a time-independent Hamiltonian, it is possible to choose the generating function as follows:
\be
W(x,\,q,\,t) = S(x,\,q) + Et
\ee
where $E$ is a constant and $q$ is a parameter that will be identified with the final position of the system. 
Then Eq. (\ref{eqn: H-J}) reduces to the time-independent Hamilton-Jacobi equation
\be \label{eqn: time independent H-J}
\frac{1}{2m}|\nabla S|^{2} + V(x)=E \, .
\ee

Once a complete solution $S(x\,,q)$ of the Hamilton-Jacobi equation is determined, it can be shown that 
$$
A^{2}= \left|\det\left(\frac{\partial^{2} S}{\partial x^{j}\,\partial q^{k}}\right) \right| \, ,
$$ 
is a solution of Eq. (\ref{eqn pseudo-continuity equation}).
Indeed, a complete solution of the Hamilton-Jacobi equation is a function $S(x,q)$, with parametric dependence on the second factor $q$, that would define a diffeomorphism  $\mathrm{d}S\colon\mathbb{R}^{n}\times\mathbb{R}^{n}\rightarrow T^{*}\mathbb{R}^{n}$ given by\footnote{It is possible to generalize this construction to an  arbitrary configuration manifold $Q$ (see for instance \cite{carinena_gracia_martinez_munoz-lecanda_roman-roy-geometric_hamilton-jacobi_theory}).
In such case, $\mathrm{d}S$ will be a diffeomorphism only on open submanifolds of $Q\times Q$ and $T^{*}Q$.}:
\be
\mathrm{d}S (x,q) := (x, \mathrm{d}_qS(x)) \, ,
\ee
or, in local coordinates, $p_k = \partial S/\partial q^k (x,q)$, by means of which we can replace initial position $x$ and initial momentum $p$, with initial position $x$ and final position $q$.
Furthermore, we can define a symplectic structure $\omega$ on $\mathbb{R}^{n}\times\mathbb{R}^{n}$ as follows:

\be
\omega:=\left(\mathrm{d}S\right)^{*}\omega_{0}=\mathrm{d}\left((\mathrm{d}S)^{*}\theta_{0}\right)=\frac{\partial S}{\partial x^{j}\partial q^{k}}\,\mathrm{d}x^{j}\wedge\mathrm{d}q^{k}\,,
\ee
where $\theta_{0}=p_{j}\,\mathrm{d}q^{j}$ is the the tautological one form on $T^{*}\mathbb{R}^{n}$, and $\omega_{0}=\mathrm{d}\theta_{0}=\mathrm{d}p_{j}\wedge\mathrm{d}q^{j}$ is the canonical symplectic structure on $T^{*}\mathbb{R}^{n}$ \cite{abraham_marsden-foundations_of_mechanics}.
Consider the Hamiltonian function $H=\frac{1}{2m} p^2+ V(x)$ on $T^{*}\mathbb{R}^{n}$ and its associated Hamiltonian vector field $X_{H}$ defined by the condition $i_{X_{H}}\omega_{0}=\mathrm{d}H$, then, since $\mathrm{d}S$ is a diffeomorphism, it is possible to define the vector field 
\be
\tilde{X}_{H}=((\mathrm{d}S)^{-1})_{*}X_{H}\,
\ee
which is the image under the push-forward of the diffeomorphism $(\mathrm{d}S)^{-1}$ of the Hamiltonian vector field $X_{H}$.
The vector field $\tilde{X}$ will be the Hamiltonian vector field, with respect to the symplectic structure $\omega$ on the manifold $\mathbb{R}^{n}\times \mathbb{R}^{n}$, of the Hamiltonian:

\be
\tilde{H}=\mathrm{d}S^{*}(H) \, .
\ee
Since the new variables $q^{j}$ will be constants of the motion, we will have that: 
\be
\tilde{X}_{H} = \frac{\partial S}{\partial x^{j}}\frac{\partial}{\partial x^{j}} \, .
\ee

In order to solve Eq. (\ref{eqn pseudo-continuity equation}) it is better to rewrite it in a more useful form. 
If we multiply both sides of the equation by $2A$ we get the following expression:
\be \label{eqn: divergence equation}
\frac{\partial A^{2}}{\partial t} + \nabla\cdot \left( A^{2}\frac{\nabla S}{m} \right) = 0 \, .
\ee

If we consider a time-independent solution $A$, the equation above  becomes:
\be \label{eqn: delta equation}
\nabla \cdot\left( A^{2}\frac{\nabla S}{m} \right) = 0 \, .
\ee

In order to exhibit an explicit solution of this equation let us notice that if a vector field $X$ preserves the volume form  $f\Omega$, then the vector field $fX$ preserves the volume $\Omega$ (see \cite{carinena_gracia_marmo_martinez_munoz-lecanda-a_quantum_route_to_hamilton-jacobi_equation:comments_and_remarks} for a proof). 
Since: 

\be
L_{X}\Omega = (\mathrm{div}X) \Omega\,,
\ee
the previous result tells us that if the divergence of $X$ with respect to the volume form $f\Omega$ is zero, then the divergence of $fX$ with respect to the volume form $\Omega$ is zero. 
We already know that the vector field $\tilde{X}_{H}$ is Hamiltonian with respect to the symplectic form $\omega$; therefore, it preserves the volume form:
$$
f\Omega = \det \left( \frac{\partial^2 S}{\partial x^j \partial q^k} \right) dx^{1}\wedge dq^{1}\wedge \cdots \wedge dx^{n}\wedge dq^{n} \, .
$$ 
It then follows that the vector field $f\tilde{X}_{H}$ preserves the volume form $\Omega$ on $\mathbb{R}^{n}\times \mathbb{R}^{n}$, and thus we get that: 

\be 
\nabla \cdot\left( \det \left( \frac{\partial^{2}S}{\partial x^{j} \partial q^{k}} \right) \nabla S \right) = 0\,.
\ee 
Consequently, that a stationary solution of Eq. (\ref{eqn pseudo-continuity equation}) is given by: 

\be 
A^2 = \det \left (\frac{\partial^{2}S}{\partial x^{j} \partial q^{k}} \right) \, .
\ee

The existence of a complete solution for the Hamilton-Jacobi equation, however, implies the system to be completely integrable, that is, there must be $n$ independent constants of the motion in involution, which is a very special situation.

It should be noticed however that, formally, quantum Hamiltonian systems are completely integrable in some properly defined setting (\cite{clemente-gallardo_marmo-towards_a_definition_of_quantum_integrability, cirelli_pizzocchero-on_the_integrability_of_quantum_mechanics_as_an_infinite_dimensional_system}).
Thus, when the resulting classical limit will not be integrable, one may think that the deleted ``quantum potential term'' in Eq. (\ref{eqn: Q-H-J}) may be responsible for the loss of integrability.

It is clear that (\ref{eqn: H-J}) is a non-linear equation, and thus, the classical limit, understood as $\hbar\rightarrow 0$,  has destroyed the linearity of Schr\"{o}dinger equation.
Looking at this situation from the opposite point of view, we could say that the non-linear Hamilton-Jacobi equation becomes linear.
Moreover, when we add the amplitude $A$ and unfold the resulting system into a Hilbert-space setting, it becomes  completely integrable.
Therefore, it could be tempting to say that the ``quantization'' procedure can be thought of as a possible linearization procedure for a first order partial differential equation.

Hence, it is readily seen that the classical limit of the Schr\"{o}dinger equation highly depends on the chosen wavefunction $\psi(x\,,t)$, since an arbitrary wavefunction $\psi=A\,\mathrm{e}^{-\frac{\imath}{\hbar} W}$ needs not to be such that $\frac{\hbar^{2}}{2m}\frac{\Delta A}{A}\approx 0$.
Accordingly, it seems that the information on the ``classical limit'' of the theory is not contained in the whole Hilbert space $\mathcal{H}$, but in families of suitably-defined states.
This idea of considering subsystems is at the basis of the so-called reduction procedures, which have been fruitfully employed in the Hamiltonian description of dynamical systems. 
Indeed, a common situation arising in reduction procedures is precisely the generation of nonlinear dynamics starting from linear ones (see for instance \cite{carinena_ibort_marmo_morandi-geometry_from_dynamics_classical_and_quantum}, Ch. 7.1-2). 
In the following, we will try to set the stage for a reduction-like analysis of quantum dynamical maps, leading to  dynamical maps on submanifolds of quantum states which could be interpreted as classical-like ones. 
In addition it will be shown that such reduction procedure does not depend on performing a $\hbar \to 0$ limit, thus separating the emergence of classical-like structures on subsystems of  quantum states from any \textit{ad hoc} zero limit of the Planck's constant (see for instance the discussion on the computation of the $\hbar \to 0$ limit of Gr\"onewal's kernel in \cite[Sect. 2.4]{ibort_manko_marmo_simoni_stornaiolo_ventriglia-the_quantum_to_classical_transition:contraction_of_associative_products}).

In Classical Mechanics, (pure) states of a system are described as points of a suitable manifold $M$, usually a phase-space.
The observables of the theory are described as a certain class of real-valued functions on $M$.
Of course, if $M$ is a smooth manifold, the observables are described by real-valued smooth functions.
The dynamical evolution is described using a one-parameter group $\gamma_{t}$ of transformations of $M$ in itself.
If $M$ is a smooth manifold, and $\gamma_{t}$ is smooth, then there is a (complete) vector field $\Gamma$ generating $\gamma_{t}$.
The fact that $\gamma_{t}$ is a one-parameter group is associated with the fact that we want the evolution of the system to be completely determined once the initial state $m\in M$ is specified.

In Quantum Mechanics, it is possible to immerse a manifold $M$ in the Hilbert space $\mathcal{H}$ of a physical system, for instance, by means of the so-called generalized coherent states.
In this way, to every $m\in M$ there corresponds a normalized vector $|m\rangle\in\mathcal{H}$, and it is possible to define real-valued functions on $M$ starting with quantum observables (described by self-adjoint operators), and viceversa.
From this point of view, generalized coherent states can be thought of as a double-way bridge between Classical and Quantum Mechanics, for, on the one hand, they can be used as a tool to achieve the quantization of a given classical system, and, on the other hand, they can be used as a tool to ``dequantize'' a given quantum system \cite{ali_antoine_gazeau-coherent_states_wavelets_and_their_geeralizations}.

We want to pursue this last perspective by implementing a reduction-like procedure of quantum dynamical maps using generalized coherent states.
Specifically, we ask if it is possible to immerse $M$ in $\mathcal{H}$ in such a way that a given quantum unitary evolution on $\mathcal{H}$ defines a one-parameter group of transformations of $M$ in itself.
If so, we can ask if there is a classical-like interpretation for the points of $M$, and thus, for the dynamical system on it arising from the reduction of the quantum dynamical map.
In this sense we interpret the resulting dynamica map as being classical-like.
Of course, a complete answer to this question is not easy to give, and thus we limit ourselves to a preliminary discussion in which the conceptual aspects of this project are outlined and the well-known example of the canonical coherent states for the quantum harmonic oscillator is reformulated accordingly.


\section{Weyl systems and generalized coherent states}

In the spirit of Dirac correspondence principle, classical Poisson-Brackets on functions on a phase space are replaced by commutators among linear operators on a Hilbert space. 
In the case of canonical commutation relations (CCRs) $[\mathbf{Q}\,,\mathbf{P}]=\imath\hbar\mathbb{I}$, at least one of the linear operators representing positions $\mathbf{Q}$ and momenta $\mathbf{P}$ must be an unbounded operator, leading to problems related to the domain of definition for the CCRs. 
To handle this problem Weyl proposed to formulate CCRs in terms of group elements rather than algebra generators \cite{weyl-Quantenmechanik_und_gruppentheorie}, \cite{wigner-on_the_quantum_correction_for_thermodynamic_equilibrium}.
Specifically, let $(V, \omega)$ be a symplectic Abelian vector group of finite dimension $2n$, that is, a vector space $V$ endowed with a non-degenerate antisymmetric bilinear form $\omega$ (symplectic form) invariant under the action of the vector group.  Then, in Weyl's approach the CCRs are replaced with a projective unitary representation $\mathbf{U}$ of the symplectic Abelian vector group $V$, i.e., for any $v \in V$, $\mathbf{U}(v)$ is a unitary operator on a Hilbert space $\mathcal{H}$ such that:

\be\label{eqn: weyl CCRs}
\mathbf{U}(v)\mathbf{U}(w)\mathbf{U}(v)^{\dagger}\mathbf{U}(w)^{\dagger}= \mathrm{e}^{\imath\,\omega(v,w)}\,.
\ee
By selecting a Lagrangian subspace $X \subset V$, i.e., a maximal isotropic subspace, the unitary operators $\mathbf{U}(v)$ corresponding to an irreducible representation of $V$ can be realized as von Neumann's irreducible representation on the Hilbert space $\mathcal{L}^2(X,\mathrm{d}\mu)$ of square integrable functions $\psi$ on $X$ with respect to the Lebesgue measure:

\be 
(\mathbf{U}(v)\psi)(x) = (\mathbf{U}(x,\alpha)\psi)(x) = \mathrm{e}^{\imath \alpha\cdot x} \psi(x+q)\,.
\ee    
where the symplectic vector space $V$ is naturally identified with $X \oplus X^* \cong T^*X$ and vectors $v\in V$ can be written as pairs $(x,\alpha)$ with $x \in X$ and $\alpha \in X^*$.
It is well-known that the generators $\mathbf{Q}$ of the subgroup $\mathbf{U}(q\,,0)$ and the generators $\mathbf{P}$ of the subgroup $\mathbf{U}(0\,,\alpha)$ satisfy the CCRs on an appropriate domain \cite{reed_simon-methods_of_modern_mathematical_physics_I_functional_analysis, esposito_marmo_sudarshan-from_classical_to_quantum_mechanics}.

Weyl's idea can be generalized to the so-called quantizer-dequantizer formalism \cite{manko_manko_marmo-alternative_commutation_relations_star_product_and_tomography}, in which projective representations of groups are replaced by two maps $U,D$, called quantizer and dequantizer respectively.
We start with a measure space $(M\,, \mu)$, for instance a topological space with a Borelian measure, and a Hilbert space $\mathcal{H}$ with its associated spaces $\mathcal{L}(\mathcal{H})$ and $\mathcal{U}(\mathcal{H})$ of linear and unitary operators respectively.
We consider two maps $U, D\colon M\rightarrow \mathcal{U}(\mathcal{H})$, by means of which we can associate a unitary operator $U(m)$, or $D(m)$, to any point $m\in M$.
The map $U$ allows us to build operators starting with functions on $M$, that is, given a function $f$ in $M$ we define the linear operator:
\be\label{Af}
\mathbf{A}_{f}:=\int_{M}f(m)\,U(m)\mathrm{d}\mu(m)\, ,
\ee
with $\mathbf{A}_{f}$ acting on the vector $|\psi\rangle\in \mathcal{H}$ as 
$$
\mathbf{A}_{f} |\psi \rangle :=\int_{M}f(m)(\,U(m)  |\psi \rangle ) \, \mathrm{d}\mu(m)\, .
$$
Thus $|\psi \rangle$ will be in the domain of $\mathbf{A}_{f}$ if $|| \mathbf{A}_{f} |\psi \rangle || < \infty$.   This will be achieved if the map $U$ is strongly continuous, that is for any $|\psi \rangle\in  \mathcal{H}$, the map $x \mapsto U(x) |\psi \rangle$ is continuous, and $f \in \mathcal{L}^1(M, d\mu )$.  Notice that in such case (notice that the map $x \mapsto U(x) |\psi \rangle$ is not only continuous but bounded $||U(x) |\psi \rangle|| \leq ||U(x) || ||\psi || = ||\psi||$): 
$$
|| \mathbf{A}_{f} |\psi \rangle || \leq || f ||_{\mathcal{L}^1} || \psi ||\, ,
$$
and the operator $\mathbf{A}_{f}$ is bounded.   More general measurable maps $f$ will lead to unbounded operators $\mathbf{A}_{f}$.
Analogously, starting with $D$ and a linear operator $\mathbf{A}$ we can build a function $f_{\mathbf{A}}$:

\be\label{fA}
f_{\mathbf{A}}(m):= \mathrm{Tr}\, (\mathbf{A}\,D^{\dagger}(m))\,.
\ee
Clearly, whenever $\mathbf{A}$ is an unbounded operator, a careful analysis is needed in order to be sure that the trace in the definition of $f_{\mathbf{A}}$ makes sense.

If the maps $U,D$ are such that:

\be
\int_{M}\, \mathrm{Tr} \, (D^{\dagger}(m)\,U(m'))\,f(m')\,\mathrm{d}\mu(m')=f(m)\,,
\ee
for any test function $f$ on $M$, that is 
\be\label{biorthogonality}
\mathrm{Tr} \, (D^{\dagger}(m)\,U(m')) = \delta (m,m') \, ,
\ee 
in the sense of distributions, then if $D(m)$ is strongly continuous too, it is readily seen that on test functions:
\be\label{fAf=f}
f_{\mathbf{A}_{f}}(m)= \mathrm{Tr} \, (D^{\dagger}(m)\,\mathbf{A}_{f})=\int_{M}f(m') \mathrm{Tr}\, (D^{\dagger}(m)\,U(m'))\,\mathrm{d}\mu(m')=f(m)\, ,
\ee
and if we assume that the map $f_{\mathbf{A}}$ is integrable in $M$, then the correspondence $\mathbf{A} \mapsto f_\mathbf{A}$ is a left-inverse to the correspondence $f \mapsto \mathbf{A}_f$. 

Fixing a fiducial normalized state $|0\rangle$ in the Hilbert space $\mathcal{H}$,  we can use the map $U$ to immerse $M$ in the Hilbert space $\mathcal{H}$ rather than in the unitary group $\mathcal{U}(\mathcal{H})$ by means of the map $m\in M \mapsto |m\rangle \in \mathcal{H}$ given by:
\be
|m\rangle:= U(m)|0\rangle\, . 
\ee
In general, we can immerse a classical-like manifold $M$ in the Hilbert space $\mathcal{H}$ by means of an injective immersion $i \colon M \to\mathcal{H}$ with no reference to the unitary group $\mathcal{U}(\mathcal{H})$.
The two most common features required for this map are weak continuity, that is, the map $m\mapsto \langle\psi|m\rangle$ is continuous for all $|\psi\rangle\in\mathcal{H}$, and the completeness condition:
\be
\int_{M}\,|m\rangle\langle m|\,\mathrm{d}\mu(m)=\mathbb{I}\,.
\ee
Given an orthonormal basis $\{|k\rangle\}$ of $\mathcal{H}$, we have that:
\be
|m\rangle=\sum_{k}\psi_{k}(m)|k\rangle\,,
\ee
and the completeness condition implies that the set of functions $\psi_{k}(m)$ form an orthonormal set in the Hilbert space $\mathcal{L}^{2}(M\,,\mathrm{d}\mu)$.  We will call a set of states $|m\rangle$ satisfying these properties generalized coherent states and the triple $(M, U, D)$ a quantizer-dequantizer scheme or a generalized Weyl system.

\begin{remark}
This immersion procedure is very similar to what is done in information geometry where a statistical model $\mathcal{M}$ is immersed in the statistical manifold $\mathcal{P}(X)$ of probability distributions on a measure space $X$ (see \cite{amari_nagaoka-methods_of_information_geometry}, \cite{amari-information_geometry_and_its_application} and references therein).
Indeed, we can pullback the Hermitean tensor\footnote{Note that this tensor is not defined on the null vector. 
Indeed, it is the pullback to $\mathcal{H}_{0}=\mathcal{H}-\{\mathbf{0}\}$ of an Hermitean tensor on the complex projective space $\mathcal{P}(\mathcal{H})$.}:

\be\label{eqn: hermitean tensor on the Hilbert space}
\mathfrak{h}:=\frac{\langle\mathrm{d}\psi|\mathrm{d}\psi\rangle}{\langle\psi|\psi\rangle} - \frac{\langle\mathrm{d}\psi|\psi\rangle\langle\psi|\mathrm{d}\psi\rangle}{\langle\psi|\psi\rangle^{2}}
\ee
on $M$ to obtain a Riemannian and a (pre)symplectic tensor (the real and the immaginary part of the pullback tensor).
The Riemannian tensor defined on $M$ in this way can be thought of as the Quantum Fisher-Rao metric (\cite{facchi_kulkarni_manko_marmo_sudarshan_ventriglia-classical_and_quantum_fisher_information_in_the_geometrical_formulation_of_quantum_mechanics}).
\end{remark}

Analogously to what has been done with the maps $U,D$ in Eq. (\ref{Af}) and Eq. (\ref{fA}), we may use the parametrized family of states $|m\rangle$ to build linear operators starting with functions (Notice that in this situation the operator valued function $m \mapsto |m\rangle\langle m|$ is strongly continuous):

\be
f\mapsto \mathbf{A}_{f}:=\int_{M} \, f(m)\,|m\rangle\langle m|\,\mathrm{d}\mu(m)\,,
\ee
and viceversa:

\be
\mathbf{A}\mapsto f_{\mathbf{A}}(m):=\langle m|\mathbf{A}|m\rangle\,.
\ee
If the analogue of the biorthogonality condition Eq. (\ref{biorthogonality})
is satisfied, that is:
$$
\mathrm{Tr} \, (|m\rangle\langle m| |m'\rangle\langle m'|) = \delta (m,m') \, ,
$$
then relation Eq. (\ref{fAf=f}) holds and $f_{\mathbf{A}_f} = f$ if $f$ is integrable.

Notice that $\mathbf{A}_{f+g}=\mathbf{A}_{f} + \mathbf{A}_{g}$, and $f_{\mathbf{A}+\mathbf{B}}=f_{\mathbf{A}} + f_{\mathbf{B}}$.
Of course, if $\mathbf{A}$ is unbounded, we must be sure that the vectors $|m\rangle$ lie in its domain in order for $f_{\mathbf{A}}(m)$ to make sense.
In addition, we note that $f$ is real valued if and only if  $\mathbf{A}_{f}$ is symmetric.

The correspondence $\mathbf{A}\mapsto f_{\mathbf{A}}$ allows us to use the Lie and Jordan products on self-adjoint linear operators to define a symmetric and a skew-symmetric  product on real-valued functions $f_{\mathbf{A}}$.
Indeed, let $\odot$ denote the Jordan product:

\be
\mathbf{A}\odot\mathbf{B}:=\frac{1}{2}\left(\mathbf{A}\mathbf{B} + \mathbf{B}\mathbf{A}\right)\,,
\ee
and let $[[\,,]]$ denote the Lie product:

\be
\left[\left[\mathbf{A}\,,\mathbf{B}\right]\right]:=-\frac{\imath}{\hbar}\left[\mathbf{A}\,,\mathbf{B}\right]\,,
\ee
on pairs of self-adjoint operators.
Then, we define the brackets of the corresponding functions:

\be
\left(f_{\mathbf{A}}\,,f_{\mathbf{B}}\right):=f_{\mathbf{A}\odot\mathbf{B}}\,, \qquad
\left\{f_{\mathbf{A}}\,,f_{\mathbf{B}}\right\}:=f_{\left[\left[\mathbf{A}\,,\mathbf{B}\right]\right]}\,.
\ee
If we can find $n = \dim(M)$ linear operators $\mathbf{A}_{1}\,,...\,,\mathbf{A}_n$ such that:
\be
\mathrm{d}f_{\mathbf{A}_{1}}(m)\wedge\mathrm{d}f_{\mathbf{A}_{2}}(m)\wedge...\wedge\mathrm{d}f_{\mathbf{A}_n}(m)\neq0 \;\;\;\;\;\forall m\in M\,,
\ee
then $\left\{\mathrm{d}f_{\mathbf{A}_{1}}(m)\,,...\,,\mathrm{d}f_{\mathbf{A}_n}(m)\right\}$ is a basis of $T^{*}_{m}M$ for all $m\in M$ and we can write:
\be\label{G}
G\left(\mathrm{d}f_{\mathbf{A}_{j}}\,,\mathrm{d}f_{\mathbf{A}_{k}}\right):=\left(f_{\mathbf{A}_{j}}\,,f_{\mathbf{A}_{k}}\right)\,,
\ee
\be\label{Lambda}
\Lambda\left(\mathrm{d}f_{\mathbf{A}_{j}}\,,\mathrm{d}f_{\mathbf{A}_{k}}\right):=\left\{f_{\mathbf{A}_{j}}\,,f_{\mathbf{A}_{k}}\right\}\,.
\ee
Given $f_{1},f_{2}$ arbitrary (real-valued)  smooth functions on $M$ we can expand their differentials in terms of the chosen basis:
\be
\mathrm{d}f_{1}=\alpha_{1}^{j}\,\mathrm{d}f_{\mathbf{A}_{j}}\,,\;\;\;\;\;\mathrm{d}f_{2}=\alpha_{2}^{j}\,\mathrm{d}f_{\mathbf{A}_{j}}\,,
\ee
and thus we can define the following $(2,0)$ tensors $G$ and $\Lambda$:
\be\label{GG}
G\left(\mathrm{d}f_{1}\,,\mathrm{d}f_{2}\right):=\alpha^{j}_{1}\,\alpha^{k}_{2}\,G\left(\mathrm{d}f_{\mathbf{A}_{j}}\,,\mathrm{d}f_{\mathbf{A}_{k}}\right)\,,
\ee
\be\label{LambdaLambda}
\Lambda\left(\mathrm{d}f_{1}\,,\mathrm{d}f_{2}\right):=\alpha^{j}_{1}\,\alpha^{k}_{2}\,\Lambda\left(\mathrm{d}f_{\mathbf{A}_{j}}\,,\mathrm{d}f_{\mathbf{A}_{k}}\right)\,,
\ee
where the summation on repeated indices is understood.
Notice that, the linear extension of \ref{G} and \ref{Lambda} according to \ref{GG} and \ref{LambdaLambda} does not agree, in general, with the brackets among linear operators, that is,  once a choice of $\left\{\mathrm{d}f_{\mathbf{A}_{1}}(m)\,,...\,,\mathrm{d}f_{\mathbf{A}_n}(m)\right\}$ is made for all $m\in M$, it could happen that there are linear operators $\mathbf{B},\mathbf{C}$ such that:
\be\label{eqn: inequalities}
G\left(\mathrm{d}f_{\mathbf{B}}\,,\mathrm{d}f_{\mathbf{C}}\right)\neq f_{\mathbf{B}\odot\mathbf{C}} \, , \qquad  
\Lambda\left(\mathrm{d}f_{\mathbf{B}}\,,\mathrm{d}f_{\mathbf{C}}\right)\neq f_{[[\mathbf{B}\,,\mathbf{C}]]} \, .
\ee
When the submanifold $M$ is considered to be a constraint manifold, we would have a situation similar to the one considered by Dirac when dealing with constraints (\cite{dirac-lectures_on_quantum_mechanics}).
The written equalities should be understood as weak equalities in the sense of Dirac, therefore our remark follows.
In Section \ref{Section: Dynamical maps in the quantizer-dequantizer formalism} (remark \ref{remark: inequalities}), when dealing with the coherent states, we will give an explicit example where this situation is actually realized.


\section{Dynamical maps in the quantizer-dequantizer formalism}\label{Section: Dynamical maps in the quantizer-dequantizer formalism}

\subsection{Invariant sets of generalized coherent states}
Up to now we have been interested in a kinematical description of our system with no attention to the dynamical aspect of the theory which, in Quantum Mechanics, is encoded in a strongly continuous one-parameter group $\mathbf{U}_{t}$ of unitary operators on the Hilbert space of the system.
Now, we want to understand if a quantum dynamical map $t \mapsto \mathbf{U}_{t}$ induces a flow $\gamma_{t}$ on a classical-like manifold $M$ of generalized coherent states.
If so, we could interpret $\gamma_{t}$ as a classical-like dynamical flow on $M$ representing the quantum evolution $\mathbf{U}_t$.

The generalized reduction procedure principle (as discussed for instance in \cite{carinena_ibort_marmo_morandi-geometry_from_dynamics_classical_and_quantum}, Ch. 7), states that a necessary condition for $\mathbf{U}_{t}$ to induce a dynamical map $\gamma_{t}$ on $M$ is the invariance of the range $\Sigma = i(M)$ of the immersion $i$ as a subset of $\mathcal{H}$, with respect to $\mathbf{U}_{t}$, that is $\mathbf{U}_{t}(\Sigma)\subseteq\Sigma$.
In the generalized coherent states setting this means that, for all $t \in \mathbb{R}$, $m\in M$ there exists $m_{t}\in M$ such that:

\be
\mathbf{U}_{t}|m\rangle=|m_{t}\rangle\,.
\ee
Then, the induced flow $\gamma_{t}$ in $M$ is defined as follows:
\be
m\mapsto \gamma_{t}(m):=m_{t}\,.
\ee
In general, this reduction procedure would give rise to a non-linear flow on $M$, although we started with the dynamical map $\mathbf{U}_{t}$ given by linear operators.

Notice that if $\gamma_{t}$ exists, it must be a one-parameter group of transformations of $M$.
Indeed, being $\mathbf{U}_{t}\,\mathbf{U}_{-t}=\mathbb{I}$, we naturally have that $\gamma_{-t}$ is the inverse map of $\gamma_{t}$, and viceversa.
This fact has an immediate  consequence, that is, the set $M$ can not be interpreted as a classical-like configuration space, but rather it should be thought of as a classical-like space of states, i.e., a phase-space, representing a subset of quantum states.
This follows from the fact that in general the dynamics induced on configuration space are not one-parameter groups of transformations, but just projections of flows on phase spaces.
For example, the motion of a particle in Classical Mechanics calls for the introduction of the cotangent bundle of its configuration space in order to describe its dynamics by means of a vector field, which, in turn, gives rise to a one-parameter groups of transformations.

Hence, if we have an invariant set of generalized coherent states $|m\rangle$, $m\in M$, with respect to the quantum dynamical map $\mathbf{U}_t$, it reduces to a one-parameter group of transformations $\gamma_{t}$ of $M$, and if $M$ is a smooth manifold and the maps $\gamma_t$ are smooth, then the resulting vector field $\Gamma$ describing the dynamics on $M$ must be complete. 
On the contrary, dynamical vector fields in classical Lagrangian and Hamiltonian Mechanics are often not complete because of the presence of singularities.
Notice that further reductions of the dynamical system $(M\,,\Gamma)$ can happen, as it is often the case, for instance if the original quantum system has a symmetry group and such group acts on $M$ equivariantly.

Hamilton-Jacobi theory can help us in finding examples where such a reduction is possible.
Indeed when a dynamical system is completely integrable, it admits a description in terms of action-angle variables and the corresponding dynamical flow is a one parameter group of transformations.

\subsection{Invariance and complete integrability}

We will now argue that under the conditions above, when $M$ is a boundaryless differentiable manifold, $\gamma_{t}$ should actually be the flow of a dynamical system which is completely integrable.
Suppose that  $\mathbf{H}$ is a self-adjoint operator generating the quantum dynamical map $\mathbf{U}_{t}=\mathrm{e}^{-\imath\frac{\mathbf{H}t}{\hbar}}$.  For simplicity it will be assumed that the Hamiltonian operator $\mathbf{H}$ has a purely discrete spectrum  and that the flow $\gamma_{t}$ exists.  Furthermore, again just for the sake of simplicity, we shall assume that the spectrum $\sigma(\mathbf{H})$ of the Hamiltonian operator is non-degenerate, however, the extension of the argument to the degenerate case presents no conceptual difficulties.

Let $\{|k\rangle\}$ denote the basis of normalized eigenvectors of $\mathbf{H}$, and $\mathbf{E}_{k}=|k\rangle\langle k|$ the orthogonal projector associated to the eigenvector $|k\rangle$.
Without loss of generality, we can assume that the immersion $i\colon M\rightarrow\mathcal{H}$ is such that the states $|m\rangle$ are in the domain of $\mathbf{H}$ for all $m\in M$ if $\mathbf{H}$ is an unbounded operator.
Concretely, this amounts to say that the coefficients $\psi_{k}(m)=\langle k|m\rangle$ are such that:

\be
f_{\mathbf{H}^{2}}(m)=\langle m|\mathbf{H}^{2}|m\rangle= \sum_{k} E^{2}_{k}|\psi_{k}(m)|^{2}<+\infty\;\;\forall m\in M\,,
\ee
where $E_{k}$ denotes the $k$-th eigenvalue of $\mathbf{H}$.

Since $[\mathbf{E}_{k}\,,\mathbf{U}_{t}]=0$ for all $k$, we have that $f_{\mathbf{E}_{k}}$ is a constant of the motion for $\gamma_{t}$:

\be
f_{\mathbf{E}_{k}}(\gamma_{t}(m))=\langle m|\mathbf{U}_{t}^{\dagger}\,\mathbf{E}_{k}\,\mathbf{U}_{t}|m\rangle=\langle m|\mathbf{E}_{k}|m\rangle=f_{\mathbf{E}_{k}}(m)\,.
\ee
Of course, these functions will not be all functionally independent, however, since we have an infinite number of them, it could be possible that we are able to find a subset  $f_{\mathbf{E}_{k_{1}}}\,,...\,,f_{\mathbf{E}_{k_{N}}}$ of constants of the motion where $N$ is such that the system is completely integrable. 
More generally, let us consider the algebra $\mathcal{C}$ generated by the functions $f_{\mathbf{E}_{k}}$.
Then the vector field $\Gamma$ whose flow is given by $\gamma_t$ will project to the space defined by the algebra $\mathcal{C}$.    
Now, if the skew-symmetric tensor $\Lambda$ defined before, see Eq. (\ref{Lambda}), is non-degenerate and its inverse $\omega$ defines a closed 2-form, and if the algebra $\mathcal{C}$ has $\frac{1}{2}\dim M$ independent generators, the system $\Gamma$ will be completely integrable (see  for instance \cite{carinena_ibort_marmo_morandi-geometry_from_dynamics_classical_and_quantum}, Ch. 8).

In order to make this construction more concrete, we will now look at the paradigmatic example of coherent states, namely, the canonical coherent states of the Harmonic oscillator.
In this case, the standard creation and annihilation operators $\mathbf{a}^{+}$ and $\mathbf{a}$ are such that the Hamiltonian operator is $\mathbf{H}=\hbar\omega(\mathbf{a}^{+}\mathbf{a} + \frac{1}{2}\mathbb{I})$, and its spectrum is given by $\{\hbar\omega(n+\frac{1}{2})\}$.
The canonical coherent states are given by the map:

\be
z\mapsto|z\rangle=\mathrm{e}^{z\mathbf{a} - \overline{z}\mathbf{a}^{\dagger}}\,|0\rangle=\mathrm{e}^{-\frac{|z|^{2}}{2}}\sum_{n=0}^{+\infty}\,\frac{z^{n}}{\sqrt{n!}}\,|n\rangle\,,
\ee
where $z\in M=\mathbb{C}$, and $|n\rangle$ is the n-th eigenvector of $\mathbf{H}$.
An explicit calculation shows that:

$$
\mathbf{U}_{t}|z\rangle=\mathrm{e}^{-\imath \frac{\omega t}{2}}\,\mathrm{e}^{-\frac{|z|^{2}}{2}}\sum_{n=0}^{+\infty}\,\frac{z^{n}}{\sqrt{n!}}\mathrm{e}^{-\imath n\omega t}\,|n\rangle=
$$
\be
=\mathrm{e}^{-\imath \frac{\omega t}{2}}\,\mathrm{e}^{-\frac{|z|^{2}}{2}}\sum_{n=0}^{+\infty}\,\frac{(z\mathrm{e}^{-\imath \omega t})^{n}}{\sqrt{n!}}\,|n\rangle=\mathrm{e}^{-\imath \frac{\omega t}{2}}\,|z\mathrm{e}^{-\imath \omega t}\rangle\,.
\ee
Since $\mathrm{e}^{-\imath \frac{\omega t}{2}}$ is an overall phase factor, it bears no physical relevance, and we can dispose of it.
Equivalently, we could have started considering the Hamiltonian operator $\hbar\omega\mathbf{a}^{\dagger}\mathbf{a}$, and we would have obtained directly:

\be
\mathbf{U}_{t}|z\rangle=|z\mathrm{e}^{-\imath \omega t}\rangle\,,
\ee
without the overall phase factor.

\begin{remark}
The appearence of the overall phase factor $\mathrm{e}^{-\imath\frac{\omega t}{2}}$ suggests that a more geometrical formulation of the reduction procedure outlined here should be performed considering the immersion of $M$ in the complex projective space $\mathcal{P}(\mathcal{H})$ rather than in the Hilbert space $\mathcal{H}$.
Indeed, $\mathcal{P}(\mathcal{H})$ is precisely the space of pure states of Quantum Mechanics, and, according to \cite{cirelli_pizzocchero-on_the_integrability_of_quantum_mechanics_as_an_infinite_dimensional_system}, there is an infinite-dimensional formulation of complete integrability which applies directly to unitary evolutions on $\mathcal{P}(\mathcal{H})$.
\end{remark}

Considering $\hbar\omega\mathbf{a}^{\dagger}\mathbf{a}$ as our Hamiltonian operator, we see that the dynamical evolution of a canonical coherent state is again a canonical coherent state, therefore, the quantum dynamical map associated to the Hamiltonian operator $\mathbf{H}$ gives rise to a classical-like dynamical map $\gamma_{t}$.
Since $|z\rangle$ is in the domain of the Hamiltonian for all $z\in \mathbb{C}$,
the one-parameter group $\gamma_{t}$ is differentiable, and thus, there is a complete vector field $\Gamma$ generating it.

Writing\footnote{Notice that $x$ and $p$ are dimensionless.} $z=x+\imath p$, with $x,p\in\mathbb{R}$, we immediately see that:

\be
\gamma_{t}(x\,,p)=\left(x\cos(\omega t) + p\sin(\omega t)\,,p\cos(\omega t) - x\sin(\omega t)\right)\,,
\ee
and it is clear that this is nothing but the dynamical flow of the harmonic oscillator on $M=\mathbb{C}\cong\mathbb{R}^{2}$ which is a completely integrable system.

The functions $f_{\mathbf{E}_{k}}$ are constants of the motion for $\gamma_{t}$:

\be
f_{\mathbf{E}_{k}}(x\,,p)=\hbar\omega \mathrm{e}^{-(x^{2} + p^{2})}\,\frac{(x^{2}+p^{2})^{k}}{(k-1)!}\,.
\ee
Furthermore, the function $f_{\mathbf{H}}$ is well defined for all $z\in \mathbb{C}$ and reads:

\be
f_{\mathbf{H}}(x\,,p)=\hbar\omega(x^{2} + p^{2})\,.
\ee
This is precisely the functional form of the Hamiltonian function for the classical harmonic oscillator, and, of course, it is a constant of the motion for $\gamma_{t}$.
Being $\dim(M)=2$, there can not be two (or more) functionally independent constants of the motion, and in fact, we have $\mathrm{d}f_{\mathbf{H}}\wedge\mathrm{d}f_{\mathbf{E}_{j}}=\mathrm{d}f_{\mathbf{H}}\wedge\mathrm{d}f_{\mathbf{E}_{k}}=\mathrm{d}f_{\mathbf{E}_{k}}\wedge\mathrm{d}f_{\mathbf{E}_{j}}=0$ for all $k,j$.

We will now see that, in this case, $\gamma_{t}$ is the flow of the Hamiltonian vector field $\Gamma$ associated with $f_{\mathbf{H}}$ by means of the symplectic structure $\Omega$ on $M=\mathbb{C}\cong\mathbb{R}^{2}$ constructed as follows.
Consider $\mathbf{X}=\sqrt{\frac{\hbar}{2m\omega}}(\mathbf{a}^{\dagger} + \mathbf{a})$ and $\mathbf{P}=\imath\sqrt{\frac{\hbar m \omega}{2}}(\mathbf{a}^{\dagger} - \mathbf{a})$, a direct calculation shows that the real-valued functions $f_{\mathbf{X}}$ and $f_{\mathbf{P}}$ are:

\be
f_{\mathbf{X}}(x\,,p)=\sqrt{\frac{2\hbar}{m\omega}}\,x\,,\;\;\;\;f_{\mathbf{P}}(x\,,p)=\sqrt{2\hbar m \omega}\,p\,,
\ee
and are functionally independent on all $\mathbb{C}$.
The commutation relations associated with $\mathbf{X}$ and $\mathbf{P}$ are $[\mathbf{X}\,,\mathbf{P}]=\imath\hbar\mathbb{I}$.
Obviously, thess commutation relations do not make sense on the whole Hilbert space $\mathcal{H}$ because $\mathbf{X}$ and $\mathbf{P}$ are unbounded operators, however, they do make sense, weakly, on the set of coherent states, that is, $\langle z|[\mathbf{X}\,,\mathbf{P}]|z\rangle$ is well defined for all $z$.
This means that we can calculate $\{f_{\mathbf{X}}\,,f_{\mathbf{P}}\}$:

\be
\{f_{\mathbf{X}}\,,f_{\mathbf{P}}\}=\Lambda\left(\mathrm{d}f_{\mathbf{X}}\,,\mathrm{d}f_{\mathbf{P}}\right)=1\,.
\ee
From this it follows that we can define the following antisymmetric contravariant tensor on $\mathbb{C}\cong\mathbb{R}^{2}$:

\be\label{eqn: poisson tensor for coherent states}
\Lambda=\frac{1}{\hbar}\,\frac{\partial}{\partial x}\wedge\frac{\partial }{\partial p}\,.
\ee
It is clear that this is an invertible Poisson tensor.
Its inverse $\Omega$ is a symplectic form, and reads:

\be
\Omega=\hbar\,\mathrm{d}p\wedge\mathrm{d}x\,.
\ee
A straightforward  calculation shows that $\Gamma=\Lambda(\mathrm{d}f_{\mathbf{H}}\,,\cdot)$ is indeed the vector field generating $\gamma_{t}$.
Note that the antisymmetric part $\Omega'$ of the pullback to $M=\mathbb{C}\cong\mathbb{R}^{2}$ of the Hermitean tensor $\mathfrak{h}$ (see Eq. \ref{eqn: hermitean tensor on the Hilbert space}) is:

\be
\Omega'=\mathrm{d}p\wedge\mathrm{d}x\,,
\ee
and thus $\Omega=\hbar\Omega'$.

\begin{remark}\label{remark: inequalities}
Going back to (\ref{eqn: inequalities}), we will now provide an explicit realization of the situation considered there.
At this purpose, let us consider the self-adjoint operators $\mathbf{A}=\frac{1}{2}\left(|1\rangle\langle0| + |0\rangle\langle1|\right)$ and $\mathbf{B}=\frac{\imath}{2}\left(|1\rangle\langle 0| - |0\rangle\langle 1|\right)$.
The associated functions are $f_{\mathbf{A}}=\mathrm{e}^{-(x^{2} + p^{2})}\,x$ and $f_{\mathbf{B}}=\mathrm{e}^{-(x^{2} + p^{2})}\,p$.
A direct calculation shows that $[[\mathbf{A}\,,\mathbf{B}]]=\frac{1}{2\hbar}\,\left(|0\rangle\langle 0| - |1\rangle\langle 1|\right)$, and thus
\be\label{eqn: serve nel remark 1}
f_{[[\mathbf{A}\,,\mathbf{B}]]}=\frac{\mathrm{e}^{-(x^{2} + p^{2})}}{2\hbar}\,\left(1 - x^{2} -p^{2}\right)\,.
\ee
However, if we compute the bracket between $f_{\mathbf{A}}$ and $f_{\mathbf{B}}$ using the Poisson tensor (\ref{eqn: poisson tensor for coherent states}) we obtain:
\be
\Lambda(\mathrm{d}f_{\mathbf{A}}\,,\mathrm{d}f_{\mathbf{B}})=\frac{\mathrm{e}^{-2(x^{2} + p^{2})}}{2\hbar}\,\left(1- 2x^{2} - 2p^{2}\right)\,,
\ee
which is different from (\ref{eqn: serve nel remark 1}).
An analogous result holds for $G(\mathrm{d}f_{\mathbf{A}}\,,\mathrm{d}f_{\mathbf{B}})$.
\end{remark}
 
It is interesting to note that the same classical-like dynamical map can be found starting with a Hamiltonian $\mathbf{H}$ having a non-degenerate, purely discrete spectrum with polynomial growth $\sigma(\mathbf{H})=\{\sum_{j=0}^{N}\hbar\omega\,\epsilon_{j} n^{j}\}$, that is:

\be
\mathbf{H}=\sum_{n=0}^{+\infty}\,E(n)\,|n\rangle\langle n|\,,
\ee
with $E(n)=\sum_{j=0}^{N}\hbar\omega\,\epsilon_{j} n^{j}$.
To see this, let us start with the polar form of the immersion map defining the canonical coherent states:

\be
|z\rangle=\mathrm{e}^{-\frac{\rho}{2}}\sum_{n=0}^{+\infty}\,\frac{\rho^{\frac{n}{2}}}{\sqrt{n!}}\mathrm{e}^{\imath n\varphi}\,|n\rangle\,,
\ee
where $z=\sqrt{\rho}\,\mathrm{e}^{\imath \varphi}$, and deform it as follows\footnote{A slightly more general set of coherent states of this form were investigated in \cite{gazeau_klauder-coherent_states_for_systems_with_discrete_and_continuous_spectrum}.}:

\be
|z\rangle=\mathrm{e}^{-\frac{\rho}{2}}\sum_{n=0}^{+\infty}\,\frac{\rho^{\frac{n}{2}}}{\sqrt{n!}}\mathrm{e}^{\imath \left(\sum_{j=0}^{N}\epsilon_{j}n^{j}\right)\varphi}\,|n\rangle\,.
\ee
Note that, unlike the case of canonical coherent states, this immersion presents a discontinuity at $z=0$.
Accordingly, we will consider $M=\mathbb{C}_{0}\cong\mathbb{R}^{2}-\{(0\,,0)\}$.

A straightforward calculation shows that:

\be
\mathbf{U}_{t}|z\rangle=\mathrm{e}^{-\frac{\rho}{2}}\sum_{n=0}^{+\infty}\,\frac{\rho^{\frac{n}{2}}}{\sqrt{n!}}\mathrm{e}^{\imath \left(\sum_{j=0}^{N}\epsilon_{j}n^{j}\right)(\varphi-\omega t)}\,|n\rangle\equiv|z_{t}\rangle\,,
\ee
which means that the set of coherent states is invariant with respect to the quantum dynamical map generated by $\mathbf{H}$.
Writing $z=x+\imath p$, it follows that:

\be
\gamma_{t}(x\,,p)=\left(x\cos(\omega t) + p\sin(\omega t)\,,p\cos(\omega t) - x\sin(\omega t)\right)\,,
\ee
which is again the dynamical flow of the harmonic oscillator (on $M\cong\mathbb{R}^{2}-\{(0\,,0)\}$).

The function $f_{\mathbf{H}}$ reads:

$$
f_{\mathbf{H}}(x\,,p) = \langle z|\mathbf{H}|z\rangle=\hbar\omega\,\mathrm{e}^{-\rho}\sum_{n=0}^{+\infty} \frac{\rho^{n}}{n!} \left(\sum_{j=0}^{N}\epsilon_{j} n^{j}\right)=
$$
\be
=\hbar\omega\,\sum_{j=0}^{N}\,\epsilon_{j}\,T_{j}(\rho)=\hbar\omega\,\sum_{j=0}^{N}\,\epsilon_{j}\,T_{j}(x^{2}+ p^{2})\,,
\ee
where $T_{j}(\rho)$ is the $j$-th Touchard polynomial\footnote{We recall that the j-th Touchard polynomal is  defined as follows:
$$
T_{j}(x):=\mathrm{e}^{-x}\,\sum_{k=0}^{+\infty}\,\frac{x^{k}\,k^{j}}{k!}\,.
$$
}.
The interesting fact is that the antisymmetric part $\Omega$ of the pullback to $M\cong\mathbb{R}^{2}-\{(0\,,0)\}$ of the Hermitean tensor $\mathfrak{h}$ (see Eq. (\ref{eqn: hermitean tensor on the Hilbert space})) becomes:

$$
\Omega=\left(\sum_{j=0}^{N}\,\epsilon_{j}\,\frac{\partial}{\partial \rho}T_{j}(\rho)\right)\,\mathrm{d}\rho\wedge\mathrm{d}\varphi=
$$
\be
=\frac{1}{x^{2} + p^{2}}\left(\sum_{j=0}^{N}\,\epsilon_{j}\left(x\frac{\partial}{\partial x}T_{j}(x^{2}+ p^{2}) + p\frac{\partial}{\partial p}T_{j}(x^{2}+ p^{2})\right)\right)\,\mathrm{d}x\wedge\mathrm{d}p\,,
\ee
and a direct calculation shows that the dynamical vector field $\Gamma$ generating $\gamma_{t}$ is the Hamiltonian vector field associated with $f_{\mathbf{H}}$ by means of $\hbar\,\Omega$:

\be
\hbar\,\Omega\left(\Gamma\,,\cdot\right)=\mathrm{d}f_{\mathbf{H}}\,.
\ee

From this, we conclude that, in order to guarantee that the immersed manifold $\Sigma(M)$ is invariant with respect to the different quantum dynamical maps arising from the choiche of a specific Hamiltonian $\mathbf{H}$, we have to carefully select the immersions of $M\cong\mathbb{R}^{2}-\{(0\,,0)\}$.
Once, this is done, we always obtain the same reduced classical-like dynamical map $\gamma_{t}$.
However, the function $f_{\mathbf{H}}$ and the antisymmetric part $\Omega$ of the pull-back of $\mathfrak{h}$ change in such a way that the vector field $\Gamma$ generating $\gamma_{t}$ is Hamiltonian for $f_{\mathbf{H}}$ with respect to $\Omega$, that is, different quantum dynamical maps lead to alternative Hamiltonian description of the same classical-like dynamics.

It may be interesting to observe that the immersed manifold (standard coherent states) may also be obtained, formally, by considering the critical points of the expectation-value-function $f_{\mathbf{a}}$ of the annihilation operator $\mathbf{a}$:

\be
f_{\mathbf{a}}(\psi)=\frac{\langle \psi| \mathbf{a} \psi\rangle}{\langle\psi|\psi\rangle}\,.
\ee
As $\mathbf{a}$ is not Hermitian, the eigenvalues will be all complex numbers.
From this point of view an infinitesimal generator of the dynamics should map critical points into critical points which, from the practical point of view, is easier to check than the invariance under the one-parameter group of unitary transformations.
The fact that standard coherent states are critical points follows from direct computation and the consideration that eigenvectors of the annihilation operator are associated with eigenvalues $z$.
In general, the variation should be done with independent variations on the full Hilbert space, both for bras and kets. 
We would get the equation for the critical values which conicide with the definition of critical points.
See \cite{esposito_marmo_sudarshan-from_classical_to_quantum_mechanics} or \cite{esposito_marmo_miele_sudarshan-advanced_concepts_in_quantum_mechanics}, where various definitions of coherent states are offered. 
As a spin off, it follows that all the operators in the normalizer of the annihilation operator are candidate to  define dynamics which would give a classical-like dynamics.

\section{Conclusions}

In this contribution we have tried to set the stage for the analysis of the classical limit of Quantum Mechanical systems from the point of view of the theory of reduction, that is, by introducing adequate generalized coherent states that would provide a `classical-like' interpretation of the original quantum dynamics.
The starting point is the formalism of generalized coherent states by means of which a classical-like manifold $M$ can be immersed in the space of (pure) states of Quantum Mechanics.
Once a specific dynamical map $\mathbf{U}_{t}$ on the Hilbert space $\mathcal{H}$ of the quantum system is chosen, the immersed submanifold $\Sigma(M)\subset \mathcal{H}$ of quantum states can be either invariant, that is, the evolution of a state in $\Sigma(M)$ is again in $\Sigma(M)$, or not.
If it is invariant, we can define an induced dynamical map $\gamma_{t}$ on the classical-like manifold $M$.

Natural geometrical structures in the form of two $(2,0)$ contravariant tensors are naturally induced on the manifold $M$ and, under some restrictive hypothesis, the manifold $M$ can be thought to be a symplectic manifold, that is, a good model for a classical phase space.   
In this vein, it could be argued that the induced classical-like flow must exhibit a large class of constants of the motion and, eventually, be completely integrable.

\section{Acknowledgements}
 F.M.C. and F.D.C. would like to thank the warm hospitality of ICMAT were this work was started.  A.I. was partially supported by the Community of Madrid project QUITEMAD+, S2013/ICE-2801, and MINECO grant MTM2014-54692-P.   G.M. would like to acknowledge the  partial support by the ``Excelence Chair Program, Santander-UCIIIM''.

\end{document}